\documentclass[oldversion,letter]{aa}
\usepackage{times}
\usepackage{amssymb}
\usepackage{mathptmx}
\usepackage{natbib}
\bibpunct{(}{)}{;}{a}{}{,}  
\usepackage{graphicx}
\defcitealias{nesvadba20a}{N20}
\usepackage{url}
\usepackage{txfonts}

\begin{document}
\title{Gas, dust, and star formation in the positive AGN feedback candidate 4C~41.17 at z=3.8 
  \thanks{Based on data obtained with the
IRAM NOEMA interferometer through program S19DA.}}
\author{N.~P.~H.~Nesvadba\thanks{email:
nicole.nesvadba@oca.eu}\inst{1}\and G.~V.~Bicknell\inst{2}\and D.~Mukherjee\inst{3}\and A.~Y.~Wagner\inst{4}}
\institute{
  Universit\'e de la C\^ote d'Azur, Observatoire de la C\^ote d'Azur, CNRS, Laboratoire Lagrange, Bd de l'Observatoire, CS 34229, 06304 Nice cedex 4, France
  \and
  Research School of Astronomy and Astrophysics, The Australian National University, Canberra, ACT 2611, Australia
  \and
  Inter-University Centre for Astronomy and Astrophysics, Post Bag 4, Pune - 411007, India
  \and
  University of Tsukuba, Center for Computational Sciences, Tennodai 1-1-1, 305-0006, Tsukuba, Ibaraki, Japan}
\authorrunning{Nesvadba et al.}
\titlerunning{Gas, dust, and star formation in 4C~41.17}

\date{Received  / Accepted }

\abstract{
  We present new, spatially resolved [CI]1--0, [CI]2--1,
  CO(7--6), and dust continuum observations of 4C~41.17 at
  $z=3.8$. This is one of the best-studied radio galaxies in this
  epoch and is arguably the best candidate of jet-triggered star
  formation at high redshift currently known in the
  literature. 4C~41.17 shows a narrow ridge of dust continuum
  extending over 15~kpc near the radio jet axis. Line emission is
  found within the galaxy in the region with signatures of positive
  feedback.  Using the [CI]1--0 line as a molecular gas tracer, and
  multifrequency observations of the far-infrared dust heated by star
  formation, we find a total gas mass of $7.6\times 10^{10}$
  M$_{\odot}$, which is somewhat greater than that previously found
  from CO(4--3). The gas mass surface density of $10^3$ M$_{\odot}$
  yr$^{-1}$ pc$^{-2}$ and the star formation rate surface density of
  10~M$_{\odot}$ yr$^{-1}$ kpc$^{-2}$ were derived over the
  12~kpc$\times$8~kpc area, where signatures of positive feedback have
  previously been found. These densities are comparable to those in
  other populations of massive, dusty star-forming galaxies in this
  redshift range, suggesting that the jet does not currently enhance
  the efficiency with which stars form from the gas. This is
  consistent with expectations from simulations, whereby radio jets
  may facilitate the onset of star formation in galaxies without
  boosting its efficiency over longer timescales, in particular after
  the jet has broken out of the interstellar medium, as is the case in
  4C~41.17.}


\maketitle

\section{Introduction} 
\label{sec:introduction}

It is now widely recognized that powerful radio jets ejected by
powerful active galactic nuclei (AGN) can have a major impact on the
evolution of their host galaxies by regulating their gas content and
interstellar medium conditions, thereby potentially facilitating the
formation of new stars. Outflows of molecular, atomic, and warm
ionized gas found in numerous galaxies are our prime observational
evidence of feedback today.

Other potential mechanisms, by which the injection of AGN energy may
regulate star formation in the host galaxies, are much less
explored. This includes the enhancement of gas turbulence and pressure
within the interstellar gas and star-forming clouds. A pressure
increase on embedded gas clouds may temporarily facilitate star
formation \citep[``positive feedback'';
  e.g.,][]{fragile04,wagner12,silk13,fragile17}. The existence of such
positive feedback has been expected theoretically for several decades,
but observational evidence is hard to collect, and is only found for
very few individual targets in the nearby and distant universe
\citep[][]{croft06,salome15,salome17}.  \citet[][]{maiolino17} suggest
that velocity offsets found in a nearby ultraluminous infrared galaxy
(ULIRG) may also point to positive feedback associated with
radio-quiet quasar activity.

An interesting question is whether positive feedback can boost the
efficiency with which stars form from molecular gas beyond the levels
found in intensely star-forming galaxies, as postulated by some models
of galaxy evolution \citep[][]{gaibler12, silk13, mukherjee18}. Shocks
driven into the gas by the radio jet may compress the denser regions
of these clouds and lead to star formation from gas that was
previously marginally Jeans-stable \citep[e.g.,][]{wagner12,
  fragile17}. The rare candidates at low redshift with signatures of
positive feedback often seem to fall below the relationship between
molecular gas mass and star formation rate density, suggesting these
galaxies form stars at lower efficiency than those without AGN
\citep[][]{salome17}. However, our clearest examples of positive
feedback in the nearby Universe are either dwarf galaxies \citep[e.g.,
  Minkowski's Object, 3C~285, ][]{salome15} or the outskirts of
massive early-type galaxies \citep[e.g., Cen~A, ][]{salome17}. These
examples have low gas-mass surface densities compared to actively
star-forming galaxies, and therefore they do not allow us to infer
whether or not AGN may boost star formation in environments dominated
by dense molecular gas. \citet{mukherjee18} found from relativistic
hydrodynamic simulations that the efficiency of star formation in such
galaxies likely depends strongly on the detailed properties of the gas
clouds, jet power, and interaction geometry, making it very difficult
to predict theoretically whether the net effect of the radio jet on
star formation is positive or negative.

\begin{figure*}
  \centering
  \includegraphics[width=0.4\textwidth]{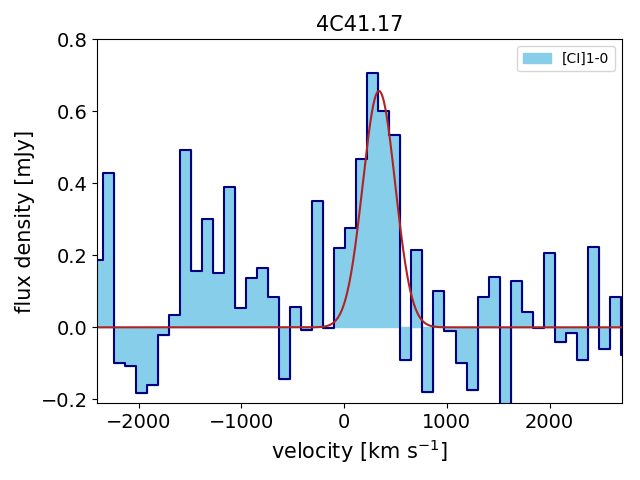}
  \includegraphics[width=0.4\textwidth]{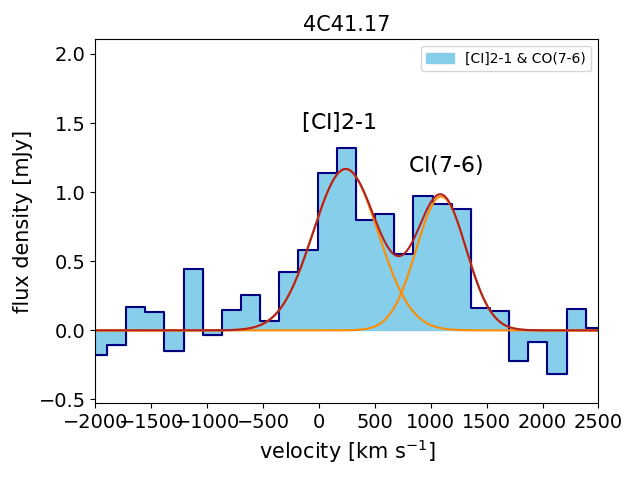}
  \caption{\label{fig:spec} Integrated spectra of [CI]1--0 {\it
      (left)}, and the [CI]2--1 and CO(7--6) lines {\it (right)}. The red
    and orange lines show the total fit to the line profile and the fits
    to the individual components, respectively.}
\end{figure*}

To test whether positive feedback may enhance the star formation
efficiency in massive galaxies in the early Universe, we observed
the powerful radio galaxy 4C~41.17 at $z=3.792$ \citep[][]{chambers90}
with the NOEMA interferometer of the Institut de Radioastronomie
Millimetrique (IRAM), probing the [CI] lines of atomic carbon and the
dust continuum at $\sim$500~GHz rest-frame frequency. We observed [CI]
rather than CO because this line is the more reliable tracer of molecular
gas mass at high redshift, it is optically thin, and is less affected by gas
excitation effects than the more commonly observed mid-J CO lines
\cite[e.g.,][]{papadopoulos04,valentino18,nesvadba19}. 4C~41.17 is
arguably our best example of positive AGN feedback in the early
Universe, as shown by the alignment of extended, unpolarized
Ly$\alpha$ emission with the radio jet axis \citep[][]{dey97},
strongly kinematically perturbed gas \citep[][]{steinbring14}, and
asymmetric absorption-emission spectra typical of very young stellar
populations \citep[P Cygni profiles;][]{dey97}. \citet{bicknell00}
present a detailed analytical study of the potential gas compression
by the radio jet, finding that at least parts of the extended
Ly$\alpha$ emission associated with the radio jet axis and bright
extended UV continuum may originate from star formation in gas
compressed through the passage of the radio jet. A young starburst age
of 30~Myr is also supported by multicomponent SED fitting
\citep[][]{rocca13} and is consistent with the age estimate of the
radio jet \citep[][]{bicknell00}.

In the following we describe our observations of the gas, dust,
and star formation in 4C~41.17, and discuss our results in the light
of star formation in massive galaxies at high redshift with and
without radio jets.  We use the flat $\Lambda$CDM cosmology from
\citet{planck18} with $H_0=67.4$ km s$^{-1}$ Mpc$^{-1}$,
$\Omega_M=0.315$, and $\Omega_{\Lambda}=1-\Omega_M$. At $z=3.792$ the
luminosity distance is $D_L= 34.41$~Gpc, and $7.26$~kpc are projected
onto one arcsecond.

\begin{figure*}
  \centering
  \includegraphics[width=0.4\textwidth]{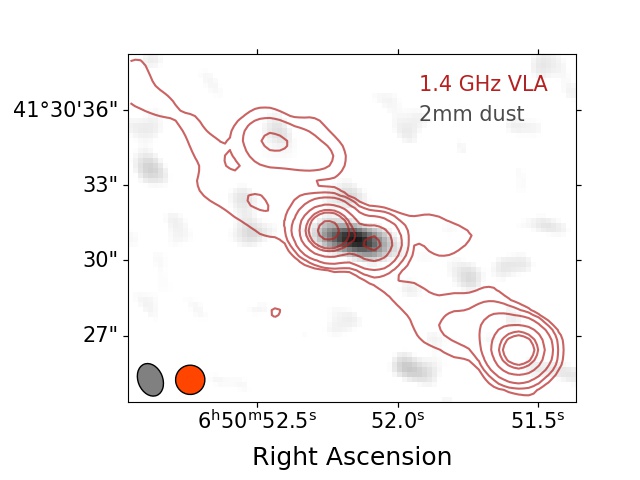}
  \includegraphics[width=0.4\textwidth]{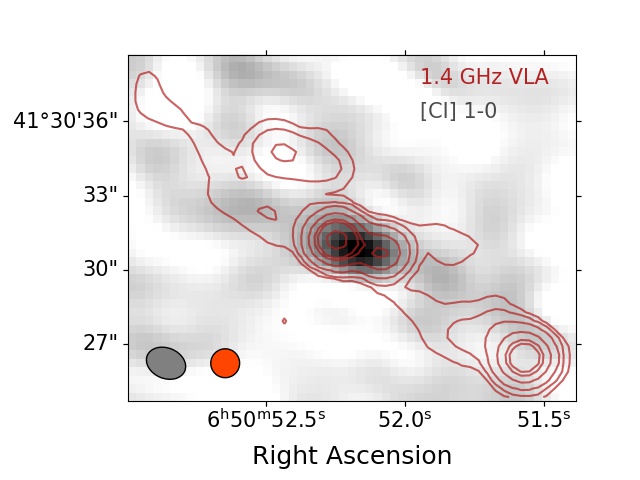}
\caption{\label{fig:morphology} Dust {\it
(left)} and [CI]1--0 morphology {\it (right)} of 4C41.17. Contours
  show the 1.4~GHz radio morphology. The radio core
  is between the two central components
  \citep[][]{bicknell00}. The beam size is shown in the lower left
  corner of each panel.}
\end{figure*} 

\section{NOEMA interferometry of 4C~41.17}
\label{sec:observations}

4C~41.17 was observed between June and December 2019 with the NOEMA
interferometer in the 2 mm and 3 mm bands with either 9 or 10 antennae
in the C configuration under average to good atmospheric
conditions. The central frequencies were $102.705$ and $168.895$~GHz,
respectively, corresponding to the observed frequencies of [CI]1--0
and [CI]2--1 at $z=3.79198$, respectively. The source was observed for
a total of 7.0 h and 5.6~h in the 2 mm and 3~mm bands,
respectively. The data were calibrated with the Continuum and
  Line Interferometer Calibration package
  CLIC\footnote{http://www.iram.fr/IRAMFR/GILDAS} and with MWC349 and
LKHA101 as flux calibrators.  Combining both polarizations leads to a
continuum root mean square (rms) of 7.3 ~$\mu$Jy bm$^{-1}$and
23.9~$\mu$Jy bm$^{-1}$ in the 3~mm and 2~mm bands, respectively,
measured over 3.8~GHz bandwidth.  We rebinned the data to a spectral
resolution of 20~MHz in both bands (corresponding to 58 km s$^{-1}$ in
band~3, and 36 km s$^{-1}$ in band~2, respectively), and obtained a
line rms of 0.1 and 0.3 mJy bm$^{-1}$ and per spectral channel,
respectively. With natural weighting, we obtain a beam size of
2.3\arcsec$\times$1.7\arcsec, and 1.4\arcsec$\times$1.0\arcsec\ at 3
mm and 2~mm, respectively.

Fig.~\ref{fig:morphology} shows the [CI] and dust morphology.  All
line emission is concentrated within an area of full width at half maximum (FWHM)
2.8\arcsec$\times$1.7\arcsec\ around the radio core for [CI]1--0, and
1.8\arcsec$\times$1.1\arcsec\ for the [CI]2--1 and CO(7--6) lines,
respectively. The major axis of the line emission exceeds the
  major axis size of the beam significantly in both bands, clearly
  showing that the emission is resolved along this axis, but not
  extended enough to neglect beam smearing. Gaussian deconvolution
  suggests major axis lengths of 1.6\arcsec\ and 1.3\arcsec\ for the
  [CI]1-0 and the [CI]2-1 and CO(7-6) lines, respectively.

The continuum, best seen in the 2~mm band, is strongly elongated and
more extended along the major axis than the line emission observed in
the same band, has an observed size of
2.4\arcsec$\times$1.1\arcsec, and a deconvolved size of
2.0\arcsec. For simplicity, and because our main goal is to rule
  out that the star formation rate surface density exceeds those
  measured in other galaxies, we adopt a common size of
  1.6\arcsec$\times$1.1\arcsec\ for the area emitting [CI]1-0 and the
  dust continuum.

The continuum major axis is roughly aligned with the radio jet
axis. Integrated continuum flux densities, which are measured after subtracting
the line emission from the data cubes, are 488$\pm$24 and
112$\pm$7~$\mu$Jy in the 2 mm and 3~mm band, respectively. Integrated
line and continuum fluxes are listed in Table~\ref{tab:fitresults}. As
already noted by \citet{debreuck05}, the nonthermal synchrotron
emission from the radio source does not significantly change these measurements.

To extract line fluxes, we fit the [CI]1--0 line with a single
Gaussian centered on a velocity of $(228\pm 21)$ km s$^{-1}$ relative to
$z=3.97197$, a FWHM line width of $(372\pm 21)$ km s$^{-1}$ and a flux
of $I_{CI10}=(0.26\pm 0.03)$ Jy~km~s$^{-1}$.

A two-component fit to the [CI]2--1 and CO(7--6) lines yields
integrated line fluxes of 0.84$\pm$0.07 Jy km s$^{-1}$ for [CI]2--1,
and 0.53$\pm$0.06 Jy km s$^{-1}$ for CO(7--6), respectively. Both
lines are significantly broader than the [CI]1--0 line: CO(7--6) has
FWHM=527$\pm$56 km s$^{-1}$, whereas [CI]2--1 has FWHM=686$\pm56$~km
s$^{-1}$. It is possible that [CI]2-1 and CO(7--6) luminosities are
enhanced near the AGN relative to [CI]1--0, since both probe
relatively high-excitation gas. The ratio $R_{21/10}=3.2$ of [CI]1--0
and [CI]2--1 flux falls near the divide between gas heated by AGN and
star formation, but into the star formation regime
\citep{meijerink07}. It is therefore possible that parts of the
      [CI]2--1 emission are from gas heated by the AGN.  It is also
      interesting to note that we do not see the same double-peaked
      line profile in CO(7--6) that \citet{debreuck05} previously
      found for CO(4--3), which may also indicate differences in gas
      excitation.

\section{Gas and star formation properties}
\label{sec:analysis}

We used equation~(4) of \citet{az13}, $M_{\rm H_2,[CI]} = 1380 \times
\frac{D_{\rm L}^2}{(1+z)}\ A_{10,-7}^{-1}\ X_{{\rm
    CI},-5}^{-1}\ Q_{10}^{-1}\ I_{\rm CI}\ [{\rm M_{\odot}}]$, where
$D_{\rm L}$ is the luminosity distance in units of Gpc, $z$ the
redshift, $I_{\rm CI}$ the integrated line flux of [CI]1--0 in Jy km
s$^{-1}$. The Einstein A coefficient, $A_{10}=7.93\times 10^{-8}$ is
given in units of $10^{-7}$ s$^{-1}$, and the carbon abundance,
  $X_{CI}=3\times 10^{-5}$ \citep{az13}, in units of $10^{-5}$.
 We find a molecular gas mass of $M_{H2}=(7.6\pm\ 1.0) \times\ 10^{10}
 M_{\odot}$, which is somewhat greater than the $5.4\times 10^{10} M_{\odot}$
 previously found by \citet{debreuck05} from CO(4--3) {with a
   7\arcsec$\times$5\arcsec\ beam; this is a sufficient to cover most of the
   surrounding Ly$\alpha$ halo. This suggests that we have not
   resolved more extended line emission associated with 4C~41.17.

We estimated the star formation rates from the dust continuum in the
NOEMA data at 2 mm and 3~mm and used previous Herschel/SPIRE and
ground-based single-dish measurements taken from the compilation of
\citet{drouart14} to determine the dust temperature, $T_D$ and the
$\beta$ parameter. We only used observations taken at $\ge350\ \mu$m,
where the AGN contamination is $<5$\% in the spectral energy distribution (SED) of dust shown by
\citet{drouart14}. This suggests $T_D=(53\pm3)$~K and $\beta=1.9$,
respectively, for a modified blackbody fit calculated in the same way
as in \citet[][]{canameras15}. This SED overpredicts the observed flux
densities by factors~2.5 and 2 in the 2 mm and 3~mm bands, respectively
(Table~\ref{tab:fitresults}). This may either be due to missing flux
in the interferometry data or additional, fainter dust emission
within the much larger beam size (10-20\arcsec) of the single-dish
observations. A potential candidate for additional far-infrared (FIR)
emission would be diffuse emission from the large,
5\arcsec$\times$7\arcsec, Ly$\alpha$ halo previously found by
\citet[][]{vanbreugel99}. In the following we use the fluxes
obtained with NOEMA, which are best matched to the [CI] emission-line
region rather than those derived from the single-dish measurements.

We adopted a modified version of the relationship of
\citet{kennicutt98b} to estimate a star formation rate from the FIR
luminosity. Using the Chabrier stellar initial mass function this
implies $SFR\ [{\rm M}_{\odot}\ {\rm yr}^{-1}] = 2.5\times
10^{-44}\ L_{FIR}\ [{\rm erg\ s}^{-1}]$. We find a total
star formation rate of $SFR=650\ {\rm M}_{\odot}$ yr$^{-1}$. This is a
factor $\sim 2$ greater than the $SFR=290\ M_{\odot}$ yr$^{-1}$ of
star formation that \citet{vanbreugel99} found for the UV continuum in
the same region, which is likely affected by dust extinction. Averaged
  over a region of
  1.6\arcsec$\times$1.1\arcsec\ (11.6~kpc$\times$8~kpc at $z=3.79$),
  this corresponds to star formation and gas mass surface densities of
  $\Sigma_{SFR} = 8.9 M_{\odot}$ yr$^{-1}$ kpc$^{-2}$ and
  1040~M$_{\odot}$ yr$^{-1}$ pc$^{-2}$, respectively. Since we
  measured and resolved [CI] and the dust continuum with similar beam
  sizes, these results are robust, even if some of the discrepancy
  between the continuum measurements with NOEMA and single-dish
  telescopes should be due to missing flux instead of additional
  astrophysical sources. 

\section{Discussion and summary}
\label{sec:summary}

Gas mass and star formation rate surface density in galaxies are
closely related; they have a near-linear relationship that holds over more
than six orders of magnitude in star formation and for molecular gas
mass surface densities $>10\ M_{\odot}$ pc$^{-2}$
\citep[e.g.,][]{kennicutt98, bigiel08}. Actively star-forming galaxies
at high redshift tend to fall above that relationship, as first noted
by \citet{daddi10} and \citet{genzel10}, indicating shorter gas
consumption times and potentially higher star formation efficiency.

Fig.~\ref{fig:sk} shows where 4C~41.17 falls in this Schmidt-Kennicutt
diagram between star formation and gas mass surface density,
  averaged over the 11.6~kpc$\times$8~kpc area previously defined in
  Section~\ref{sec:observations}, and the gas mass and
star formation rate from Section~\ref{sec:analysis}. We also add the
second galaxy with potential evidence of jet-triggered star formation
in the early Universe, PKS~0529-549 at $z=2.6$, to this analysis.
This galaxy was previously discussed by \citet{man19}, who
point out that star formation in this source seems to be more
efficient than in giant molecular clouds in the Milky Way. Both
galaxies fall well above the typical relationship found for disk
galaxies in this redshift range, which are on the main sequence of
star formation, as already noted by \citet{man19}. However, the
star formation rates do not fall above those found in other intensely
star-forming galaxies at the same cosmic epoch \citep{bothwell10}, which 
likely include the progenitors of powerful high-redshift radio
galaxies.

This result suggests that the star formation efficiency does not
exceed that found in other strongly star-forming galaxies without AGN,
at least when averaged over the $\sim 10$~kpc regions where
jet-triggered star formation was found \citep{dey97}.  We might suspect that positive and negative feedback would both be present and
partially cancel the effect on each other
\citep[][]{dugan17,mukherjee18}. \citet{steinbring14} found outflow
signatures in warm ionized gas in 4C~41.17, suggesting that negative
feedback is at work, although the low line width of [CI]1--0 does not
suggest that a large percentage of the cold neutral gas is currently
being removed. However, in situ gas removal through winds would also
lower the observed gas-mass surface densities. This would move the
galaxy to the left in Fig.~\ref{fig:sk} and artificially enhance, not
lower, the observed star formation efficiency.

A more likely explanation is that the phase of positive feedback is
very short and only triggers the onset of star formation without
boosting it over longer timescales. This is not unexpected from a
theoretical point of view: Initial hydrodynamic simulations suggested that jet-triggered star formation may be long-lived
\citep{gaibler12}, but did not take into account energy and momentum
losses through disk porosity. More recent work suggests that the
impact of the jet on the star formation drops after a short initial
``kick-off' phase \citep[][]{fragile17, mukherjee18}. This would also
be consistent with the very high star formation rates of 7300
$M_{\odot}$ yr$^{-1}$ required to form the $2.2\times 10^{11}$
M$_{\odot}$ of young stars found by \citet{rocca13} within 30 Myr, that is, an
order of magnitude higher than today. Fig.~\ref{fig:morphology} also
shows that the jet has already broken out of 4C~41.17, suggesting it
may no longer be able to sustain a high overpressure within the
gas. Jet-triggered star formation may also be limited by star
formation itself once the starburst has evolved past the first few
million years, as previously suggested by \citet{fragile17}. At 30~Myr age
\citep[][]{rocca13}, the starburst in 4C~41.17 is clearly evolved
enough to have produced the first generations of supernovae. These may
make star formation in 4C~41.17 self-regulating, as has previously
been suggested for other massive, intensely star-forming galaxies at
$z=2-3$ \citep[e.g.,][]{lehnert13}.

\begin{table}
  \centering
  \begin{tabular}{lcccc}
    \hline
    \hline  
Line   &  Velocity     & FWHM         & $I_{line}$        & $S_{cont}$ \\
       & [km s$^{-1}$] & [km s$^{-1}$] & [Jy km s$^{-1}$] & [mJy]     \\
\hline
  $[$CI$]$1--0  &  228$\pm$21& 372$\pm$21& 0.26$\pm$0.03 & 0.11$\pm$0.015\\
  $[$CI$]$2--1  & 237$\pm$52 & 686$\pm$56& 0.84$\pm$0.07 & 0.49$\pm$0.035\\
  CO(7--6)      & 95$\pm$55  & 526$\pm$56& 0.53$\pm$0.06 & 0.49$\pm$0.035\\
  \hline
  \hline
\end{tabular}
\caption{\label{tab:fitresults} Observational results. Velocities are
  given relative to $z=3.792$.}
\centering
\end{table}

To summarize, we used new [CI]1--0 and 2--1, CO, and dust
continuum interferometry obtained with IRAM/NOEMA to study star
formation in 4C~41.17 at z=3.8, which is the best example of positive feedback
from radio jets at high redshift in the literature. We estimated a
molecular gas mass of $7.6\times 10^{10}$ M$_{\odot}$ in an elongated
region where signatures of jet-triggered star formation have
previously been found. The spatially resolved gas and star formation
rate densities suggest that stars are formed at efficiencies that are
comparable to submillimeter galaxies at the same epoch. The same is
true for PKS~0529-549, the second galaxy with likely signatures of a
very young, potentially jet-triggered, stellar population.

While our results do not call into question the previous good evidence
for jet-triggered star formation in 4C~41.17 obtained in the UV and
radio, they do suggest that this phase is very short and not
sustained after the jet has broken out from the interstellar gas of
the host galaxy. After few 10~Myrs, jet-triggered star formation
proceeds with similar efficiency as in other intensely star-forming
galaxies, making the signatures of positive feedback difficult to
identify observationally. This result is also consistent with recent
observational and theoretical findings in low-redshift galaxies, which indicate that
jet-triggered star formation does not necessarily lead to a boost in
star formation efficiency, even at high gas-mass surface densities of
up to 1000 M$_{\odot}$ pc$^{-2}$.

\begin{figure}
  \includegraphics[width=0.52\textwidth]{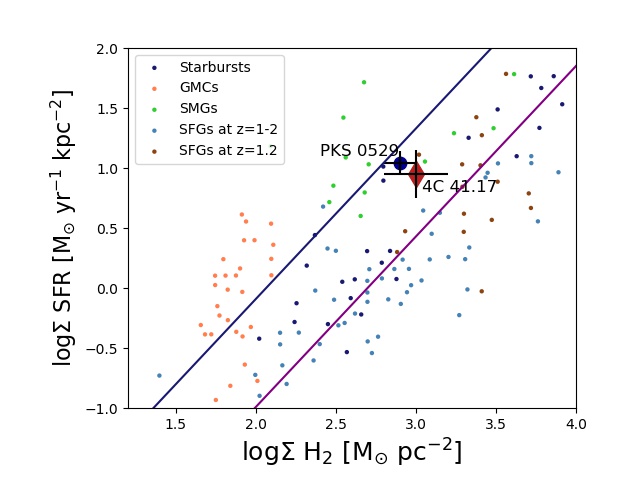}
  \caption{\label{fig:sk} Gas and star formation rate surface
    densities of 4C~41.17 (red diamond), PKS 0529-549 (blue dot), and
    several comparison samples: Giant molecular clouds in the Milky
    Way \citep[][]{evans14}, local starburst galaxies
    \citep[][]{kennicutt98b}, star-forming galaxies at $z=1-2$
    \citep[][]{tacconi08,valentino18,valentino20}, and submillimeter galaxies
    \citep[][]{bothwell10}. The solid purple and blue lines indicate the
    sequences for typical main-sequence and starburst galaxies of
    \citet{daddi10}.}
\end{figure}

\begin{acknowledgements}
We thank the referee for having gone through our paper carefully, and
for comments which helped clarify some aspects of our analysis. We
also thank the IRAM staff, in particular Charl\`ene Lef\`evre, for
taking these data, and their excellent support with the data
reduction.
\end{acknowledgements}

\bibliographystyle{aa}
\bibliography{hzrg}

\begin{thebibliography}{37}
\expandafter\ifx\csname natexlab\endcsname\relax\def\natexlab#1{#1}\fi

\bibitem[{{Alaghband-Zadeh} {et~al.}(2013){Alaghband-Zadeh}, {Chapman},
  {Swinbank}, {Smail}, {Danielson}, {Decarli}, {Ivison}, {Meijerink}, {Weiss},
  \& {van der Werf}}]{az13}
{Alaghband-Zadeh}, S., {Chapman}, S.~C., {Swinbank}, A.~M., {et~al.} 2013,
  \mnras, 435, 1493

\bibitem[{{Bicknell} {et~al.}(2000){Bicknell}, {Sutherland}, {van Breugel},
  {Dopita}, {Dey}, \& {Miley}}]{bicknell00}
{Bicknell}, G.~V., {Sutherland}, R.~S., {van Breugel}, W. J.~M., {et~al.} 2000,
  \apj, 540, 678

\bibitem[{{Bigiel} {et~al.}(2008){Bigiel}, {Leroy}, {Walter}, {Brinks}, {de
  Blok}, {Madore}, \& {Thornley}}]{bigiel08}
{Bigiel}, F., {Leroy}, A., {Walter}, F., {et~al.} 2008, \aj, 136, 2846

\bibitem[{{Bothwell} {et~al.}(2010){Bothwell}, {Chapman}, {Tacconi}, {Smail},
  {Ivison}, {Casey}, {Bertoldi}, {Beswick}, {Biggs}, {Blain}, {Cox}, {Genzel},
  {Greve}, {Kennicutt}, {Muxlow}, {Neri}, \& {Omont}}]{bothwell10}
{Bothwell}, M.~S., {Chapman}, S.~C., {Tacconi}, L., {et~al.} 2010, \mnras, 405,
  219

\bibitem[{{Canameras} {et~al.}(2015){Canameras}, {Nesvadba}, {Guery}, \&
  {others}}]{canameras15}
{Canameras}, R., {Nesvadba}, N., {Guery}, D., \& {others}. 2015, \mnras, 362,
  41

\bibitem[{{Chambers} {et~al.}(1990){Chambers}, {Miley}, \& {van
  Breugel}}]{chambers90}
{Chambers}, K.~C., {Miley}, G.~K., \& {van Breugel}, W.~J.~M. 1990, \apj, 363,
  21

\bibitem[{{Croft} {et~al.}(2006){Croft}, {van Breugel}, {de Vries}, {Dopita},
  {Martin}, {Morganti}, {Neff}, {Oosterloo}, {Schiminovich}, {Stanford}, \&
  {van Gorkom}}]{croft06}
{Croft}, S., {van Breugel}, W., {de Vries}, W., {et~al.} 2006, \apj, 647, 1040

\bibitem[{{Daddi} {et~al.}(2010){Daddi}, {Elbaz}, {Walter}, {Bournaud},
  {Salmi}, {Carilli}, {Dannerbauer}, {Dickinson}, {Monaco}, \&
  {Riechers}}]{daddi10}
{Daddi}, E., {Elbaz}, D., {Walter}, F., {et~al.} 2010, \apjl, 714, L118

\bibitem[{{De Breuck} {et~al.}(2005){De Breuck}, {Downes}, {Neri}, {van
  Breugel}, {Reuland}, {Omont}, \& {Ivison}}]{debreuck05}
{De Breuck}, C., {Downes}, D., {Neri}, R., {et~al.} 2005, \aap, 430, L1

\bibitem[{{Dey} {et~al.}(1997){Dey}, {van Breugel}, {Vacca}, \&
  {Antonucci}}]{dey97}
{Dey}, A., {van Breugel}, W., {Vacca}, W.~D., \& {Antonucci}, R. 1997, \apj,
  490, 698

\bibitem[{{Drouart} {et~al.}(2014){Drouart}, {De Breuck}, {Vernet}, {Seymour},
  {Lehnert}, {Barthel}, {Bauer}, {Ibar}, {Galametz}, {Haas}, {Hatch},
  {Mullaney}, {Nesvadba}, {Rocca-Volmerange}, {R{\"o}ttgering}, {Stern}, \&
  {Wylezalek}}]{drouart14}
{Drouart}, G., {De Breuck}, C., {Vernet}, J., {et~al.} 2014, \aap, 566, A53

\bibitem[{{Dugan} {et~al.}(2017){Dugan}, {Gaibler}, {Bieri}, {Silk}, \&
  {Rahman}}]{dugan17}
{Dugan}, Z., {Gaibler}, V., {Bieri}, R., {Silk}, J., \& {Rahman}, M. 2017,
  \apj, 839, 103

\bibitem[{{Evans} {et~al.}(2014){Evans}, {Heiderman}, \&
  {Vutisalchavakul}}]{evans14}
{Evans}, Neal~J., I., {Heiderman}, A., \& {Vutisalchavakul}, N. 2014, \apj,
  782, 114

\bibitem[{{Fragile} {et~al.}(2017){Fragile}, {Anninos}, {Croft}, {Lacy}, \&
  {Witry}}]{fragile17}
{Fragile}, P.~C., {Anninos}, P., {Croft}, S., {Lacy}, M., \& {Witry}, J. W.~L.
  2017, \apj, 850, 171

\bibitem[{{Fragile} {et~al.}(2004){Fragile}, {Murray}, {Anninos}, \& {van
  Breugel}}]{fragile04}
{Fragile}, P.~C., {Murray}, S.~D., {Anninos}, P., \& {van Breugel}, W. 2004,
  \apj, 604, 74

\bibitem[{{Gaibler} {et~al.}(2012){Gaibler}, {Khochfar}, {Krause}, \&
  {Silk}}]{gaibler12}
{Gaibler}, V., {Khochfar}, S., {Krause}, M., \& {Silk}, J. 2012, \mnras, 425,
  438

\bibitem[{{Genzel} {et~al.}(2010){Genzel}, {Tacconi}, {Gracia-Carpio},
  {Sternberg}, {Cooper}, {Shapiro}, {Bolatto}, {Bouch{\'e}}, {Bournaud},
  {Burkert}, {Combes}, {Comerford}, {Cox}, {Davis}, {Schreiber},
  {Garcia-Burillo}, {Lutz}, {Naab}, {Neri}, {Omont}, {Shapley}, \&
  {Weiner}}]{genzel10}
{Genzel}, R., {Tacconi}, L.~J., {Gracia-Carpio}, J., {et~al.} 2010, \mnras,
  407, 2091

\bibitem[{{Kennicutt}(1998{\natexlab{a}})}]{kennicutt98b}
{Kennicutt}, Robert~C., J. 1998{\natexlab{a}}, \araa, 36, 189

\bibitem[{{Kennicutt}(1998{\natexlab{b}})}]{kennicutt98}
{Kennicutt}, Robert~C., J. 1998{\natexlab{b}}, \apj, 498, 541

\bibitem[{{Lehnert} {et~al.}(2013){Lehnert}, {Le Tiran}, {Nesvadba}, {van
  Driel}, {Boulanger}, \& {Di Matteo}}]{lehnert13}
{Lehnert}, M.~D., {Le Tiran}, L., {Nesvadba}, N.~P.~H., {et~al.} 2013, \aap,
  555, A72

\bibitem[{{Maiolino} {et~al.}(2017){Maiolino}, {Russell}, {Fabian}, {Carniani},
  {Gallagher}, {Cazzoli}, {Arribas}, {Belfiore}, {Bellocchi}, {Colina},
  {Cresci}, {Ishibashi}, {Marconi}, {Mannucci}, {Oliva}, \&
  {Sturm}}]{maiolino17}
{Maiolino}, R., {Russell}, H.~R., {Fabian}, A.~C., {et~al.} 2017, \nat, 544,
  202

\bibitem[{{Man} {et~al.}(2019){Man}, {Lehnert}, {Vernet}, {De Breuck}, \&
  {Falkendal}}]{man19}
{Man}, A. W.~S., {Lehnert}, M.~D., {Vernet}, J. D.~R., {De Breuck}, C., \&
  {Falkendal}, T. 2019, \aap, 624, A81

\bibitem[{{Meijerink} {et~al.}(2007){Meijerink}, {Spaans}, \&
  {Israel}}]{meijerink07}
{Meijerink}, R., {Spaans}, M., \& {Israel}, F.~P. 2007, \aap, 461, 793

\bibitem[{{Mukherjee} {et~al.}(2018){Mukherjee}, {Bicknell}, {Wagner},
  {Sutherland}, \& {Silk}}]{mukherjee18}
{Mukherjee}, D., {Bicknell}, G.~V., {Wagner}, A. e.~Y., {Sutherland}, R.~S., \&
  {Silk}, J. 2018, \mnras, 479, 5544

\bibitem[{{Nesvadba} {et~al.}(2019){Nesvadba}, {Ca{\~n}ameras}, {Kneissl},
  {Koenig}, {Yang}, {Le Floc'h}, {Omont}, \& {Scott}}]{nesvadba19}
{Nesvadba}, N.~P.~H., {Ca{\~n}ameras}, R., {Kneissl}, R., {et~al.} 2019, \aap,
  624, A23

\bibitem[{{Papadopoulos} \& {Greve}(2004)}]{papadopoulos04}
{Papadopoulos}, P.~P. \& {Greve}, T.~R. 2004, \apjl, 615, L29

\bibitem[{{Planck Collaboration} {et~al.}(2018){Planck Collaboration},
  {Aghanim}, {Akrami}, {Ashdown}, {Aumont}, {Baccigalupi}, {Ballardini},
  {Banday}, {Barreiro}, {Bartolo}, {Basak}, {Battye}, {Benabed}, {Bernard},
  {Bersanelli}, {Bielewicz}, {Bock}, {Bond}, {Borrill}, {Bouchet}, {Boulanger},
  {Bucher}, {Burigana}, {Butler}, {Calabrese}, {Cardoso}, {Carron},
  {Challinor}, {Chiang}, {Chluba}, {Colombo}, {Combet}, {Contreras}, {Crill},
  {Cuttaia}, {de Bernardis}, {de Zotti}, {Delabrouille}, {Delouis}, {Di
  Valentino}, {Diego}, {Dor{\'e}}, {Douspis}, {Ducout}, {Dupac}, {Dusini},
  {Efstathiou}, {Elsner}, {En{\ss}lin}, {Eriksen}, {Fantaye}, {Farhang},
  {Fergusson}, {Fernandez-Cobos}, {Finelli}, {Forastieri}, {Frailis},
  {Franceschi}, {Frolov}, {Galeotta}, {Galli}, {Ganga}, {G{\'e}nova-Santos},
  {Gerbino}, {Ghosh}, {Gonz{\'a}lez-Nuevo}, {G{\'o}rski}, {Gratton},
  {Gruppuso}, {Gudmundsson}, {Hamann}, {Handley}, {Herranz}, {Hivon}, {Huang},
  {Jaffe}, {Jones}, {Karakci}, {Keih{\"a}nen}, {Keskitalo}, {Kiiveri}, {Kim},
  {Kisner}, {Knox}, {Krachmalnicoff}, {Kunz}, {Kurki-Suonio}, {Lagache},
  {Lamarre}, {Lasenby}, {Lattanzi}, {Lawrence}, {Le Jeune}, {Lemos},
  {Lesgourgues}, {Levrier}, {Lewis}, {Liguori}, {Lilje}, {Lilley}, {Lindholm},
  {L{\'o}pez-Caniego}, {Lubin}, {Ma}, {Mac{\'{\i}}as-P{\'e}rez}, {Maggio},
  {Maino}, {Mandolesi}, {Mangilli}, {Marcos-Caballero}, {Maris}, {Martin},
  {Martinelli}, {Mart{\'{\i}}nez-Gonz{\'a}lez}, {Matarrese}, {Mauri}, {McEwen},
  {Meinhold}, {Melchiorri}, {Mennella}, {Migliaccio}, {Millea}, {Mitra},
  {Miville-Desch{\^e}nes}, {Molinari}, {Montier}, {Morgante}, {Moss}, {Natoli},
  {N{\o}rgaard-Nielsen}, {Pagano}, {Paoletti}, {Partridge}, {Patanchon},
  {Peiris}, {Perrotta}, {Pettorino}, {Piacentini}, {Polastri}, {Polenta},
  {Puget}, {Rachen}, {Reinecke}, {Remazeilles}, {Renzi}, {Rocha}, {Rosset},
  {Roudier}, {Rubi{\~n}o-Mart{\'{\i}}n}, {Ruiz-Granados}, {Salvati}, {Sandri},
  {Savelainen}, {Scott}, {Shellard}, {Sirignano}, {Sirri}, {Spencer},
  {Sunyaev}, {Suur-Uski}, {Tauber}, {Tavagnacco}, {Tenti}, {Toffolatti},
  {Tomasi}, {Trombetti}, {Valenziano}, {Valiviita}, {Van Tent}, {Vibert},
  {Vielva}, {Villa}, {Vittorio}, {Wandelt}, {Wehus}, {White}, {White},
  {Zacchei}, \& {Zonca}}]{planck18}
{Planck Collaboration}, {Aghanim}, N., {Akrami}, Y., {et~al.} 2018, ArXiv
  e-prints

\bibitem[{{Rocca-Volmerange} {et~al.}(2013){Rocca-Volmerange}, {Drouart}, {De
  Breuck}, {Vernet}, {Seymour}, {Wylezalek}, {Lehnert}, {Nesvadba}, \&
  {Fioc}}]{rocca13}
{Rocca-Volmerange}, B., {Drouart}, G., {De Breuck}, C., {et~al.} 2013, arXiv
  e-prints, arXiv:1301.1983

\bibitem[{{Salom{\'e}} {et~al.}(2015){Salom{\'e}}, {Salom{\'e}}, \&
  {Combes}}]{salome15}
{Salom{\'e}}, Q., {Salom{\'e}}, P., \& {Combes}, F. 2015, \aap, 574, A34

\bibitem[{{Salom{\'e}} {et~al.}(2017){Salom{\'e}}, {Salom{\'e}},
  {Miville-Desch{\^e}nes}, {Combes}, \& {Hamer}}]{salome17}
{Salom{\'e}}, Q., {Salom{\'e}}, P., {Miville-Desch{\^e}nes}, M.~A., {Combes},
  F., \& {Hamer}, S. 2017, \aap, 608, A98

\bibitem[{{Silk}(2013)}]{silk13}
{Silk}, J. 2013, \apj, 772, 112

\bibitem[{{Steinbring}(2014)}]{steinbring14}
{Steinbring}, E. 2014, \aj, 148, 10

\bibitem[{{Tacconi} {et~al.}(2008){Tacconi}, {Genzel}, {Smail}, {Neri},
  {Chapman}, {Ivison}, {Blain}, {Cox}, {Omont}, {Bertoldi}, {Greve},
  {F{\"o}rster Schreiber}, {Genel}, {Lutz}, {Swinbank}, {Shapley}, {Erb},
  {Cimatti}, {Daddi}, \& {Baker}}]{tacconi08}
{Tacconi}, L.~J., {Genzel}, R., {Smail}, I., {et~al.} 2008, \apj, 680, 246

\bibitem[{{Valentino} {et~al.}(2018){Valentino}, {Magdis}, {Daddi}, {Liu},
  {Aravena}, {Bournaud}, {Cibinel}, {Cormier}, {Dickinson}, {Gao}, {Jin},
  {Juneau}, {Kartaltepe}, {Lee}, {Madden}, {Puglisi}, {Sanders}, \&
  {Silverman}}]{valentino18}
{Valentino}, F., {Magdis}, G.~E., {Daddi}, E., {et~al.} 2018, \apj, 869, 27

\bibitem[{{Valentino} {et~al.}(2020){Valentino}, {Magdis}, {Daddi}, {Liu},
  {Aravena}, {Bournaud}, {Cortzen}, {Gao}, {Jin}, {Juneau}, {Kartaltepe},
  {Kokorev}, {Lee}, {Madden}, {Narayanan}, {Popping}, \&
  {Puglisi}}]{valentino20}
{Valentino}, F., {Magdis}, G.~E., {Daddi}, E., {et~al.} 2020, \apj, 890, 24

\bibitem[{{van Breugel} {et~al.}(1999){van Breugel}, {Stanford}, {Dey},
  {Miley}, {Stern}, {Spinrad}, {Graham}, \& {McCarthy}}]{vanbreugel99}
{van Breugel}, W., {Stanford}, A., {Dey}, A., {et~al.} 1999, in The Most
  Distant Radio Galaxies, ed. H.~J.~A. {R{\"o}ttgering}, P.~N. {Best}, \& M.~D.
  {Lehnert}, 49

\bibitem[{{Wagner} {et~al.}(2012){Wagner}, {Bicknell}, \& {Umemura}}]{wagner12}
{Wagner}, A.~Y., {Bicknell}, G.~V., \& {Umemura}, M. 2012, \apj, 757, 136

\end{thebibliography}

\end{document}